\documentclass[epjCONF]{svjour}
\usepackage{graphics}
\usepackage[latin1]{inputenc}
\session-title{The Space Photometry Revolution}
\begin{document}
\title{Prelude to, and Nature of the Space Photometry Revolution}
\author{Ronald L. Gilliland\inst{1}\fnmsep\thanks{\email{gillil@stsci.edu}}}
\institute{Space Telescope Science Institute and Department of Astronomy
and Astrophysics, and Center for Exoplanets and Habitable Worlds, 
The Pennsylvania State University, 525 Davey Lab, University Park,
PA 16802, USA}
\abstract{
It is now less than a decade since CoRoT initiated the space
photometry revolution with breakthrough discoveries, and five
years since {\em Kepler} started a series of similar advances.
I'll set the context for this revolution noting the status 
of asteroseismology and exoplanet discovery as it was 15-25 years ago 
in order to give perspective on why it is not mere hyperbole
to claim CoRoT and {\em Kepler} fostered a revolution in our sciences.
Primary events setting up the revolution will be recounted.
I'll continue with noting the major discoveries in hand, and 
how asteroseismology and exoplanet studies, and indeed our
approach to doing science, have been forever changed thanks
to these spectacular missions.
} 
\maketitle

\section{Introduction}
\label{intro}

I would like to start by thanking Rafa Garcia, J\'{e}r\^{o}me Ballot and the Science
Organizing Committee for inviting me to give the introductory
talk at this conference.  It is a great honor for me to do so.
Of course we're here to discuss the results that come from CoRoT
and {\em Kepler}; these were French and American missions respectively
with Annie Baglin as the PI for CoRoT and Bill Borucki as the
PI for {\em Kepler}.

I would like to thank a few individuals that I have particularly benefited
from and enjoyed working with.  Thirty-five years in the case of Tim Brown,
twenty years in the case of Bill Borucki, Dave Koch, Jon Jenkins,
J{\o}rgen Christensen-Dalsgaard, and Hans Kjeldsen and about five
years now with Bill Chaplin.

In planning my talk I decided to focus on the word Revolution
and ask whether or not that was justified in characterizing 
what has happened to our field over the last several years.
So I looked in the dictionary and found the first definition:
``The forceful overthrow of a government."  Of course that's not
what we're here to talk about.  But, it is interesting that we're
talking about French and American missions, and France and the 
United States had revolutions of this type over two centuries ago
that certainly changed our countries, and fundamentally changed 
the world as well.  

The second definition is:  ``Dramatic and wide-ranging change in the way
something works or how people view it."  And that of course does fit
what we're talking about.

To focus things a little more I found a couple of quotes from the 
well-known 20th century revolutionary, Vladimir Lenin.  First of all,
``A revolution is impossible without a revolutionary situation."
One of my goals is to set the stage by describing what the 
revolutionary situation was for our two sciences.
And secondly, ``It is impossible to predict the time and progress
of revolution, it is governed by its own more or less mysterious
laws."  What could be more mysterious than how NASA and ESA and 
other national funding agencies work?

In an alternate Universe we might well be having a meeting with
the title ``The Space Photometry Revolution", but it could almost
as easily have a subtitle of MONS Symposium 3, Eddington EASC-7
joint meeting.  {\em Kepler} and CoRoT were fortunate enough to be the
missions that were flown and ultimately worked very well.
But, there were other missions proposed along the way that were
probably of roughly equal capability, and the effort that was
spent in justifying these other missions was certainly not a waste,
and was a real positive in creating the environment, the revolutionary
situation that allowed {\em Kepler} and CoRoT to eventually succeed.
There were many other missions, e.g., STARS, PRISMA, that preceded Eddington.
Jaymie Matthews's MOST of course has been very influential, although
I think not rising to the level of truly revolutionary. 
Even missions like WIRE allowed some early results in asteroseismology
from space that were influential.

\section{Status in 1994}

I would like to begin by defining what the situation was in 1994.
That's a nice round number of 20 years ago.  It also happens that, 
at that time, there were not yet any significant results in exoplanets
or asteroseismology.  There were some exotic results like pulsar 
planets, and there were close misses for asteroseismology.
I should note that for asteroseismology I'm restricting myself
to solar type stars and red giants, so I'm not attempting to do 
justice to the fundamentally important results that came out for
white dwarfs, RoAp stars, classical pulsators by Don Kurtz, 
Don Winget, Steve Kawaler and a large number of other people
that served (during the long absence of results for solar-type stars)
to keep the field of asteroseismology moving forward.

In 1994 there was an {\em Annual Reviews} article titled ``Asteroseismology"
by Tim Brown and myself\cite{brow94}, there was a GONG (Global Oscillations
Network Group), conference titled ``GONG '94: Helio and Asteroseismology from 
the Earth and Space" and there were papers presented on hoped-for
missions:  CoRoT\cite{cata94a}, STARS\cite{frid94}, {\em Kepler}\cite{milf94},
and I also presented a paper on
Photometric Methods for Asteroseismology\cite{gill94}, discussing the
possibilities of doing this from the ground.  Several members of this
audience were involved in a study that we conducted using a network of 
4-m telescopes, and we concluded that it was almost impossible to 
do from the ground.  The ideal platform would be a 1-m class telescope
in space, and I used {\em Kepler} as the primary example of that.
Also published in 1994 were contributions from a workshop that Bill Borucki put together
called ``Astrophysical Science with a Spaceborne Photometric Telescope"\cite{gran94}
and in looking at that over 20 years later, I think it's clear that the
papers Tim Brown and I separately provided for that justifying 
asteroseismology were very well posed and provided a strong case 
for including asteroseismology on {\em Kepler}.  CoRoT was proposed as 
a concept to CNES with Claude Catala as the PI\cite{cata94b}, and FRESIP
(Frequency of Earth-Sized Inner Planets) that in later iterations would
be called {\em Kepler} was proposed for the first time at the NASA 
Discovery Mission level.  It's interesting that {\em Kepler} changed its
name over the last 20 years, but kept the science that it was planning 
to do rock-solid, while CoRoT managed to keep its name but fundamentally
changed its science.  Initially it was ``Convection and 
RoTation of Stars" and later on it was changed to be ``Transits of Planets"
as the final `T', presumably partly pragmatic to get the mission to fly,
and partly from the development in the science making that an interesting
thing to actually do with a telescope that's capable of doing
asteroseismology, but rather the inverse of {\em Kepler}.

I don't mean to imply that there wasn't much earlier activity.
For instance Douglas Gough, one of the great leaders in founding the
field of asteroseismology, published a News and Views article in Nature
in 1985 \cite{goug85} in which he referred to separate French and American papers
in 1984 claiming detections on other stars.  As was generally the
case in this time period and for a long time to come, there were 
multiple claims of detections but they all tended to go away as 
better data was collected.

\section{Progress in the years 1995 -- 2009}

Over 1995 to 1999 there of course was a major breakthrough for 
exoplanets with the detection of 51 Peg b \cite{mayo95} quickly confirmed by
other groups, and these and other groups published further 
detections within this time period.  There were, however, still
skeptics throughout this time period who said these weren't 
exoplanets.  For instance some said that they were just {\em msini} detections and if
{\em sini} were unfavorable they might not be planetary masses.
The exoplanet skeptics were silenced very effectively when 
transits were detected on HD 209458 in 1999 by two independent
groups led by Tim Brown and Dave Charbonneau\cite{brow99}, and Greg Henry\cite{henr99}.
During this period there were no usable asteroseismology successes.
I say this despite several groups really trying.  There were
groups in America, groups in France, Australia was very active,
Denmark was very active; simply a very difficult thing to do 
from the ground.  The signal is very small as you know.
The efforts to advocate for a space mission that could do 
asteroseismology intensified.

There was possibly a space photometry revolution already in the 
1995-1999 period, as in 1998 and 1999 two fiercely competitive groups in
cosmology published papers showing that the Universe was accelerating.
This was obviously a revolution in our view of the Universe and since
it was partly based upon {\em HST}, time-series, space-based photometry,
it could be considered a space photometry revolution, but its not what we're
talking about.

Over 2000 to 2004 a very significant result was a space photometry
observation of HD 209458 b in transit\cite{brow01}.  Such a light curve is now 
completely routine; we've seen hundreds and thousands of these from
{\em Kepler}, many from CoRoT, MOST even occasionally, and even ground-based 
observations can nearly reach this quality now.  But in 2000 it was
anything but routine, and I like to think (and it may even be true) that
this very positive result from the {\em Hubble Space Telescope} contributed
in a significant way to {\em Kepler} and CoRoT being able to later have 
their great success.  Throughout this time period the pace of radial
velocity exoplanet detections continued to accelerate.  Asteroseismology,
finally, had a huge success\cite{bouc01}.  With Alpha Cen A there was a robust,
obvious success that was capable of supporting interpretations, and 
in this same time period there were also solid results on a giant, subgiant,
and other dwarf stars.  For Alpha Cen A, unlike previous examples where
better data resulted in evaporation of the result, here, soon enough, 
multiple groups confirmed it and improved on the results.  During this
period there were multiple space missions that were selected, and 
unfortunately dropped, but {\em Kepler} and CoRoT were both selected for 
flight, and of course as we know, (since we're
here) they were kept.

\section{Interlude -- three historical topics}

I want to take a few minutes to cover three side topics.

First is how KASC came to be.  It was always planned that for {\em Kepler},
asteroseismology was going to be a core component of the science, small
compared to exoplanets in priority, but very much part of the mission.
Tim Brown and I had submitted a budget to NASA as part of the science team support
requesting roughly \$750,000 for the full mission that would primarily
go to support a postdoctoral associate.  Tim and I always recognized
that our interest in solar-type stars and red giants was going to be 
under-supported compared to the flood of data that we expected that {\em Kepler}
would produce, and we had already started talking in a relaxed way with 
J{\o}rgen Christensen-Dalsgaard with helping out with theory and interpretation.  He of
course was very involved with our competitor, Eddington, in this period.
In October of 2003 ESA dropped Eddington.  This led to talks with J{\o}rgen
picking up quite a bit.  In April of 2004 NASA dropped all monetary 
support for asteroseismology within the Kepler Science Team.  Now, 
fortunately {\em Kepler} wasn't dropped, though it had been on the chopping block,
nor were short cadence observations dropped, though this had
been one of the things that was considered.  We were in a situation where
Tim and I had a mission that seemed to be going forward OK, but we had
no support for doing science with the data.  J{\o}rgen and Hans Kjeldsen had Danish
support for participating in Eddington, but that mission had gone away.
Thus we got together and started up KASC, and of course KASC primarily at
this point is a result of the efforts that Hans put into his masterful
planning and execution.

A second topic is how {\em Kepler} got its sharp point spread functions.
I suspect that many of you don't even think that is a true statement.
You would hear today, for instance, from {\em Kepler} project personnel in some
cases that {\em Kepler} has an intentionally soft point spread function to 
better support time series photometry.  Indeed pre-2003 our proposal
to NASA -- (emphasis in the original proposal):  ``Pointing noise
is suppressed . . . by SPREADING the PSF OVER MANY PIXELS."  We had
a requirement that less than 20\% of the flux should 
be within the central pixel, and that this should be uniform over the 
field of view.  In mid-2003 BALL Aerospace, which generally speaking did
a masterful job in developing {\em Kepler}, came to Bill Borucki and the science
team and said the ball had been dropped [pun intended] in terms of 
implementing a soft PSF.  As being built, some PSFs would have up to 68\%
of the light in the central pixel, most of them would be greater than 
30\%, and there would be severe field dependence.  This wasn't ANYTHING
like what we had asked for.  It was realized that it would take a few
million dollars to fix it, and a several month delay.  Mid-2003 wasn't 
a time to be asking for more money, since the mere existence of {\em Kepler}
at that point was still in doubt.  Emergency analyses were initiated 
within the science team to grapple with this.  Jon Jenkins and I were
tasked with looking at different aspects of the photometry given the 
sharp PSF, and people like Ted Dunham and Tim Brown were tasked with 
looking into what could be done to the optics to get us back to where we
wanted to be.  Jon and I came back after a few feverish weeks of work
and said that we were happy with the sharp PSF.  Given the exquisitely
good guiding expected for {\em Kepler} it looked like the photometric precision
wouldn't suffer.  I had looked into the astrometric aspects of being
able to eliminate background eclipsing binaries with centroiding through
transits, which was becoming a more important goal within the mission,
and found that, in fact, the sharp PSF would help quite a bit.  What
we have now is that the sharpest channels have 63\% of the light in the
central pixel, the lowest at 28\%.  Everything is much sharper than our
original desire; in some channels the PSF is severely 
under-sampled.  The good news is that it's the channels with the sharpest
PSFs that deliver the best photometry.  It's also the case that during 
the commissioning of {\em Kepler} we did everything we could to focus the 
telescope to deliver the sharpest PSFs.  We're no longer in a scenario
where we have soft PSFs; they're very sharp for {\em Kepler}.

The third side topic is meant to be humorous, and that is that American and
European sides were competitive with each other.  Nothing surprising about
that.  We wanted to do {\em the} same science, developing similar spacecraft,
and we were sometimes a bit, well not a {\em bit}, we were downright dismissive
of each other.  I played my part.  I maintained in the 90s and early
2000s that {\em Kepler} would out-perform STARS/Eddington for 
asteroseismology.  I said that with the full recognition that Eddington
was primarily trying to do asteroseismology with exoplanets also an 
important part of the mission, while {\em Kepler} was primarily trying to do
exoplanets with asteroseismology a very clear secondary, non-interfering
goal.  I still believe that would have been true, that {\em Kepler} would have
done better than Eddington.  It's of course too bad that we didn't have
both missions, I'm sure our science would be better as a result.
What's more interesting is that a paper came across my desk in July 2009,
a few months after {\em Kepler} was launched, and a couple of months after
I had seen a lot of {\em Kepler} data.  I was being asked by the editor of 
A\&A to provide a second opinion referee's report on a paper by Mosser
and Appourchaux\cite{moss09}.  I returned a report saying I thought it was a fine
paper, it should be published, but, gee, maybe the authors should be a 
a little more positive in how they describe the capabilities and goals
of {\em Kepler} for Asteroseismology.  This is what the paper, as submitted, had
to say about the {\em Kepler} capabilities: ``The {\em Kepler} mission compared to 
CoRoT will provide lower photometric performance . . . In 90 days, only 
the brightest F-type or the class IV targets will have . . . the 
large separation.  In a 4-year run, only the brightest G dwarfs will 
deliver a clean seismic signature. . . The asteroseismic goal of {\em Kepler}
is principally to derive information on stars hosting a planet, by the
determination of the large separation."  In response to my suggestion
that they should perhaps be a bit more positive it looked like one word
was changed in the paper that was ultimately published -- `lower' was
changed to `different', but clearly the gist remained the same, that 
expectations from across the Atlantic were that {\em Kepler} was not going to 
do very well.  Fortunately I think {\em Kepler} has done fine, and it's also
fortunate these individuals have become an integral part of the success
of {\em Kepler}, providing wonderful results interpreting {\em Kepler} data.

\section{The space photometry revolution begins}

Up to mid-2009 the revolution is nearly at hand.  For exoplanets things
continue to accelerate and space is continuing to gain prominence with
{\em HST}, MOST and CoRoT all providing results.  With asteroseismology radial
velocity detections improve and there is a very impressive result on
$\mu$ Ara\cite{bouc05}, the field has become a vibrant science, but without space there is only
one detection per year with great effort and resources.  CoRoT results
are published in this period on a few dwarf stars, and these are really
wonderful, but the number stays small and the quality is not all that 
much greater than what could be done from the ground, and certainly 
nothing like the quality obtained routinely for solar observations.
The promise of space photometry for both exoplanets and asteroseismology
continued to grow, and I think it is fair to say there is really a fever
pitch of anticipation.

2009 -- the middle of the year is when I would say the space photometry
revolution begins, and the start was the stunning results on red giant
oscillations from CoRoT published in papers from Joris De Ridder\cite{deri09} and
Saskia Hekker\cite{hekk09}, published in May, and accepted in June respectively.  These
results on red giants very firmly dispelled the fears of prominent 
theorists and observers as well that red giants simply might not be very
kind to us, that they might not show the types of oscillations that you could
do real science with.  The second part of the revolution occurred at
exactly the same time.  The {\em Kepler Mission} of course had very restrictive
data access in the early days.  I spent the months of May and June 2009
at NASA Ames, and I think it's a fair statement that I was the only
person in the world that had ready access to all the short cadence data
who cared about asteroseismology.  As soon as
the first 9 days of data, and then 33 days of data came down, I plotted
up the time-series, and produced power spectra.  It became very clear,
very early that F and G dwarfs as well showed pulsations very clearly.
A plot of effective noise showed that results were coming in near
optimistic limits, and perhaps even more importantly it appeared that
nature was going to be generally co-operative.  For instance the 
wonderful results from CoRoT on red giants were for rather luminous 
red giants, while with the early {\em Kepler} data we could immediately see
the oscillations all the way down the red giant branch, something that
CoRoT was not able to do.  At this point the revolution was clearly at
hand, and the community would know it whenever we managed to get the data 
distributed a little bit more. 

It's interesting to take a look at the revolution in numbers.  Here I
tabulate the number of exoplanet detections over five-year intervals, and
the number of asteroseismology detections.  The CoRoT results first
factor into the 2009 block, {\em Kepler} into the 2014.  In terms of fractions
CoRoT is small for every bin with the critical exception of
asteroseismology in 2009, where the red giants dominate.  Similarly {\em Kepler}
giants dominate the asteroseismology in the last bin.  For exoplanets it's not
clear that there ever really was a revolution in a numerical sense thanks
to space photometry.  A big boost -- 50\% -- of the final bin is a result of
{\em Kepler} and CoRoT, but things had been progressing at a geometric 
acceleration throughout the time period.  If, however, we include in 
addition the qualitative advances that I don't have time to address,
but we'll hear about throughout this meeting, then there clearly has
been a revolution in both fields due to space photometry.

\begin{table}[h]
\caption{Exoplanet and asteroseismic discoveries in 5 year intervals.}
\begin{tabular}{lrrrrr} 
\hline\noalign{\smallskip}
Category & $<$1994 & 1999 & 2004 & 2009 & 2014 \\ 
\noalign{\smallskip}\hline\noalign{\smallskip} 
Exoplanets & 4 & 25 & 120 & 265 & 1382 \\ 
Asteroseismology & 0 & 0 & 10 & 380 & 15000 \\ 
\noalign{\smallskip}\hline 
\end{tabular}
\end{table}

It's interesting to ask how space photometry has changed our sciences.
In 1994 any modestly believable result would have been wonderful.  Now
we have a stunningly large number of rock-solid detections in both fields
and a series of qualitative advances, such as discerning the rocky
and gaseous exoplanet boundary, the study of rotation and deep 
structural characteristics of red giants, and often results combining
exoplanets and asteroseismology to very strong interpretive effect.
In 1996 a paper with Geoff Marcy as first author had one co-author. 
We now sometimes have a particle physics flavor -- in 2014 we have a 
paper led by Geoff Marcy, that is Marcy plus 100 of his close colleagues.
Our approach to doing science has changed.

Closing thoughts.  In 1994 the {\em Annual Reviews} article that Tim Brown
and I wrote could only discuss the foundations of the field, and discuss
the hopes for the future, and mentioned space as something that would
really be valuable for driving the field forward.  This has been 
bookended now by an {\em Annual Reviews} report last year by Bill Chaplin and
Andrea Miglio\cite{chap13}, who note that ``we're entering a golden era for stellar
physics, and this is a result of the launch of the French-led CoRoT, and 
NASA {\em Kepler} missions".  Just as these two {\em Annual Reviews} articles nicely
bookend the prelude to the revolution, and what I what I would peg as the
time that's the end of the revolutionary period, now, Andrea will be 
providing the closing talk at the meeting.

It has been an exciting time to participate in the development of 
exoplanet and asteroseismology science and I feel particularly grateful for
having had the opportunity to work with both the {\em Hubble Space Telescope}
and {\em Kepler} throughout that period.  
Arguably I think those are the two greatest space missions that 
have existed for science to this point.  Our sciences have clearly 
undergone revolutions.  Success breeds continuation.  The future with
TESS, PLATO and other missions is very bright for decades
to come.

\end{document}